\documentclass[twoside,titlepage,final,12pt,pdftex]{article}
\usepackage{latexsym}
\usepackage{amssymb}

\usepackage[pdftex]{graphicx}
\usepackage{hyperref}

\catcode`\@=11
\def\ps@ppt{\def\@oddhead{\qquad Learning to See at the LHC \hfil \thepage\qquad}\def\@evenhead{\qquad\thepage \hfil {Chris Quigg} \qquad}
\def\@oddfoot{}\def\@evenfoot{}}    
\catcode`\@=12

\def\urll#1#2{\mbox{\href{#1}{\sf #2}}}

\makeatletter
   \renewcommand{\section}{\@startsection{section}{1}{0mm}
   {\baselineskip}%
   {\baselineskip}{\normalfont\normalsize\centering}}%
   \makeatother

\def\abs#1{\left| #1\right|}

\newcommand{\gevcc}{\ensuremath{\hbox{ GeV}\!/\!c^2}}
\newcommand{\gev}{\ensuremath{\hbox{ GeV}}}
\newcommand{\tev}{\ensuremath{\hbox{ TeV}}}

\def\ie{\hbox{\it i.e.}}	
	
\def\etal{\hbox{\it et al.}}

\def\phystoday#1#2#3#4{\frenchspacing{\it Phys. Today }{\bf #1}, #2 (\ifcase#3\or January\or 
         February\or March\or April\or May\or June\or July\or August\or 
         September\or October\or November\or December\fi, 19#4)}
\def\slashii#1{\setbox0=\hbox{$#1$}             % set a box for #1
   \dimen0=\wd0                                 % and get its size
   \setbox1=\hbox{\sl/} \dimen1=\wd1            % get size of /
   \ifdim\dimen0>\dimen1                        % #1 is bigger
      \rlap{\hbox to \dimen0{\hfil\sl/\hfil}}   % so center / in box
      #1                                        % and print #1
   \else                                        % / is bigger
      \rlap{\hbox to \dimen1{\hfil$#1$\hfil}}   % so center #1
      \hbox{\sl/}                               % and print /
   \fi}                                         %
\begin{document}
\begin{flushright}
	\textsf{FERMILAB-FN-0849-T} \\ January 12, 2010%\today
\end{flushright}
\vspace*{\stretch{1}}
%\HRule
\begin{center}
	{\LARGE  Learning to See at the Large Hadron Collider} 	\\[12mm]
	{\large Chris Quigg* } \\[5mm]
	Theoretical Physics Department \\
	Fermi National Accelerator Laboratory\\ Batavia, Illinois 60510 USA \\[4mm]
%	and\\[4mm]
	Physics Department, Technical University Munich\\ D-85748 Garching, Germany\\[4mm]
%	and\\[4mm]
	Arnold Sommerfeld Center for Theoretical Physics\\
	Ludwig-Maximilians-Universit\"{a}t M\"{u}nchen\\
D-80333 M\"{u}nchen, Germany\\[4mm]
%and\\[4mm]
Theory Group, Physics Department, CERN\\ CH-1211 Geneva 23, Switzerland\\[8mm]
	\parbox{4.in}{The staged commissioning of the Large Hadron Collider presents an opportunity to map gross features of particle production over a significant energy range. I suggest a visual tool---event displays in (pseudo)rapidity--transverse-momentum space---as a scenic route that may help sharpen intuition, identify interesting classes of events for further investigation, and test expectations about the underlying event that accompanies large-transverse-momentum phenomena.}	
\end{center}
%\HRule
\vspace*{\stretch{1.5}}
%%%%%%%%%%%%%%%%%%%%%%%%%%%%%%%
%							  %
%	\begin{center}			  %
%							  %
%							  %
%	\vspace*{\stretch{0.5}}	  %
%	\end{center}			  %
%							  %
%%%%%%%%%%%%%%%%%%%%%%%%%%%%%%%
		*\urll{mailto:quigg@fnal.gov}{E-mail:quigg@fnal.gov}
%\twocolumn
\newpage
\setlength{\parindent}{2ex}
\setlength{\parskip}{12pt}
\noindent
The first proton-proton collisions have occurred in CERN's Large Hadron Collider, at energies of $450\gev$ and $1.18\tev$ per beam, and the experimental collaborations have reported their initial looks at the data~\cite{LHCstatus}.
Early in 2010, the LHC is projected to run at $3.5\tev$ per beam, with the energy increasing later in the run to perhaps $5\tev$ per beam, or beyond.

The prime objective of the 2009--2010 run is to commission and ensure stable operation of the accelerator complex and the experiments. For the experiments, an essential task is to ``rediscover'' the standard model of particle physics, and to use familiar physics objects such as $W^\pm$, $Z^0$, $J\!/\!\psi$, $\Upsilon$, jets, $b$-hadrons, and top-quark pairs to tune detector performance. It will also be important to study the lay of the land in the new worlds that the LHC will open for exploration, even with instruments that may at first be imperfectly understood. 

What is true of the search of the agent of electroweak symmetry breaking and other marquee goals of the LHC is also true for the minimum-bias events that will dominate the early samples: \textit{We do not know what the new wave of exploration will reveal.} We need to approach the new landscape at each step in energy with an adventurous spirit buttressed by a purposeful plan for systematic observation.

The aim of this note is to call attention to two classics of the early investigation of multiparticle production and to suggest a small exercise that I believe can help develop the intuition of LHC experimenters and interested members of the broader community.

\P\ When experiments at the CERN Intersecting Storage Rings\footnote{($pp$ collisions at c.m.\ energies between $\sqrt{s} = 23$ and $63\gev$)} and in the 30-inch and 15-foot bubble chambers at Fermilab\footnote{(including $pp$ collisions at beam energies up to $400\gev$ and $\pi^\pm p$ collisions up to $200\gev$)} opened a new energy regime, Ken Wilson circulated a paper~\cite{kgw} entitled, ``Some Experiments in Multiple Production.'' The paper is worth reading today, both for the orderly approach that Wilson advocated and as a reminder of how little we knew, not so long ago. Here is a short summary of the ``experiments,'' which amounted in large measure to a catalogue of informative plots:
\begin{enumerate}
 \setlength{\itemsep}{1pt}
\item \textit{Topological cross sections:} Do multiplicity distributions exhibit a two-component structure, suggestive of diffractive plus multiperipheral production mechanisms?
\item \textit{Feynman scaling:} Is the single-particle density $\rho_1(k_z/E, k_\perp, E)$ independent of the beam energy $E$, when plotted in terms of Feynman's scaling variable $x_{\mathrm{F}} \equiv k_z/E$?
\item \textit{Factorization:} Is the single-particle density $\rho_1(k_z/E, k_\perp, E)$  in the backward (proton) hemisphere independent of the projectile, \ie, the same for $\pi p$ and $pp$ scattering?
\item \textit{$dx/x$ spectrum:} Does the single-particle density exhibit a flat plateau in the central region when plotted in terms of the rapidity (boost) variable, $y \equiv \frac{1}{2}\ln{[(k_0 + k_z)/(k_0 - k_z)]}$ ?
\item \textit{Double Pomeron exchange:} Do some events display low central multiplicity with large rapidity gaps on both ends?
\item \textit{Correlation length experiment:} Does the two-particle correlation function $C(y_1,y_2) \equiv \rho_2(y_1,y_2)-\rho_1(y_1)\rho_1(y_2)$ display short-range order, $\propto \exp(-\abs{y_1 - y_2}/L)$?
\item \textit{Factorization test with central trigger:} This is a refinement of \#3, to eliminate diffraction and isolate the multiperipheral component.
\end{enumerate}
This is not the place to review what we learned from Wilson's points of departure, either in the early days or in the later S$\bar{p}p$S and Tevatron Collider experiments. I simply remark that giving structure to the examination of high-multiplicity events set the stage for many new insights and ideas~\cite{Jacob:1973mg, Quigg:1973pd,Whitmore:1973ri,Foa:1975eu,Giacomelli:1979nu,Albajar:1989an},\footnote{For surveys of particle production at the Tevatron, see~\cite{Aaltonen:2009ne}.} and that events characteristic of the double-Pomeron topology have been observed~\cite{Aaltonen:2009kg} at the Tevatron. For an example of how much could be learned from a modest sample ($\sim 3500$ events in the 30-inch bubble chamber), see~\cite{Kafka:1975cz}. The ALICE collaboration has reported gross features of $pp$ collisions at 
the LHC injection energy of $450\gev$ per beam~\cite{ALICE:2009dt}.
 
\P\ The second provocative paper from early times that I wish to recall is James Bjorken's ``Multiparticle Processes at High Energy''~\cite{Bjorken:1971ww}, which contains a number of perceptive comments on single-particle densities and on two-particle correlations, as well as some fascinating speculations. The exploratory approach outlined there applies to the new experiments at the LHC, no less than to the experiments of the 1970s. 

While it might be tempting to surmise that multiparticle production in soft collisions constitutes settled knowledge (described, \textit{grosso modo}, as a superposition of diffractive scattering and short-range order), we have good reason to suspend judgment. At the highest energies, well into the ($\propto \ln^2{s}$?) growth of the $pp$ total cross section, long-range correlations might show themselves in new ways. Quantum Chromodynamics suggests new, modestly collective, effects such as multiple-parton interactions.\footnote{The high density of partons carrying  $p_z = 5\hbox{ to }10\gev$ may give rise to hot spots in the spacetime evolution of the collision aftermath, and thus to thermalization or other phenomena not easy to anticipate from the QCD Lagrangian.} The $\ln{s}$ expansion of the rapidity plateau softens kinematical constraints in the central region, and the sensitivity to high-multiplicity events  (or otherwise rare occurrences) of modern experiments vastly exceeds what could be seen with bubble-chamber statistics. 

Prior experience with soft production and the underlying event structure is informative, but it is not necessarily a full representation of what will happen at the higher energies we will explore at the LHC. \textit{I think it is entirely possible that a few percent of minimum-bias events recorded at energies above} $\sim 1\tev$ \textit{per beam will display an unusual structure that is not just a stretched-in-rapidity version of what the CDF and D\O\ experiments saw at the Tevatron.} Finding out about particle production at the LHC is of course important for engineering purposes, but may also teach us some new physics.

The FELIX collaboration proposed a full-acceptance
detector to make a comprehensive study of the strong interactions at the LHC. Questions posed in the FELIX physics document~\cite{Ageev:2001qv} reach beyond the realm of soft physics to encompass hard-scattering events. Many of them are apt for the detectors now taking data at the LHC.

In two recent talks~\cite{RDF:2009}, Rick Field has outlined a program of early measurements aimed at characterizing the minimum-bias events and the underlying event in hard collisions, emphasizing that these are related, but not identical, phenomena, and highlighting the importance of multiple parton interactions.

\begin{figure}[tb]
\centerline{\includegraphics[width=0.8\textwidth]{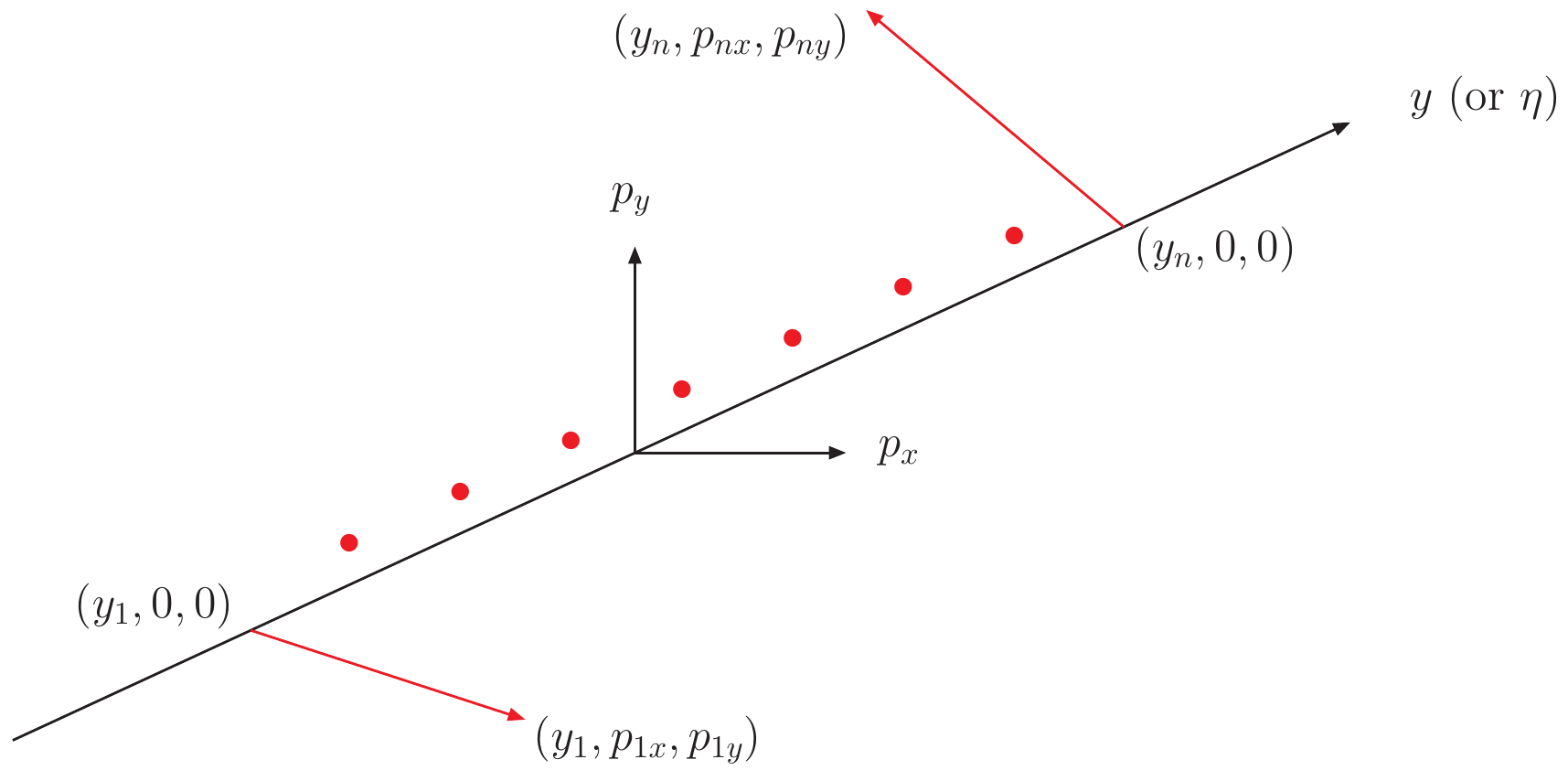}}
\caption{Schematic event display in $(y,\vec{p}_\perp)$ space. \label{fig:example}}
\end{figure}

\P\ This brings me to the specific suggestion of a visual aid that will help us learn to see at the LHC. One of the exercises suggested long ago by Bjorken~\cite{Bjorken:1971ww} was to construct three-dimensional representations of multiparticle events to draw on our human powers of visualization and pattern recognition, in the hope of recognizing important new questions. For particle production in soft collisions, it is not spatial coordinates that are most apt, but a representation in terms of (pseudo)rapidity and (two-dimensional) transverse momentum. To begin, draw a (pseudo)rapidity axis as an oblique line. Represent each track $i$ in the event by a vector drawn from $(y_i,0,0)$ to $(y_i,p_{ix},p_{iy})$, as in the example shown in Figure~\ref{fig:example} (all scales linear).

In the pre-industrial era, we drafted such event displays by hand on graph paper, which gave a certain artisanal satisfaction and promoted close scrutiny of the structure of each example as it emerged. Bjorken commented that he knew he was wiser for the experience, although he couldn't express why. My reaction was similar, once I had filled a notebook with event pictures: I couldn't tell you what I learned, but I could ask better questions of the data.
The high particle multiplicity at the LHC discourages the hands-on approach, but computers can do the drawing for us. Note that the $(y,p_\perp)$ representation is simply a curled-up vector representation of the LEGO$^{\scriptsize \textregistered}$ plot for individual tracks, with thresholds for display set as low as possible. That means that most of the technology required to generate $(y_i,p_{ix},p_{iy})$ lists is already in place.
\begin{quotation}
I encourage the LHC collaborations to produce $(y,\vec{p}_\perp)$ displays of minimum-bias events acquired during early running. Samples as small as a few hundred events would already build intuition, but I would go further. I suggest that the collaborations make available live streams of $(y,\vec{p}_\perp)$ representations, along with the online displays of events that show the structure in terms of detector elements in ordinary space. \textit{More is to be learned from the river of events than from a few examples!} Changes in event structure as a function of beam energy, or the onset of new features, might raise important questions. It is useful to color the tracks to label their charges, and to identify species where possible.\footnote{Modern computer tools make it straightforward to construct $(y,p_\perp)$ displays that can be rotated in three dimensions. The ability to manipulate events and regard them from changing perspectives can engage our perceptive powers more fully.}
\end{quotation}
Unlike the bubble chambers of old, the LHC detectors have significant threshold requirements and geometrical restrictions. But the four principal detectors have different characteristics, so it should be possible, by pooling information, to build a reasonably comprehensive picture. It is worth noting that the LHC detectors have sensitivity to neutrals that the hydrogen bubble chambers did not. {\em It would also be interesting to make streams of simulated events from the low-$p_\perp$ event generators, with and without the cuts appropriate to the four detectors.} The simulated events should also include, as separate samples, true minimum-bias events and events with hard-scattering triggers that represent the structure of underlying events. While the real comparisons will be made with statistical measures, visual examination of event streams might be an efficient way to identify the main issues.

The strong interactions comprise a richer field than the set of phenomena that we have learned to describe in terms of perturbative QCD: interesting strong-interaction processes are not merely reactions involving a few jets. The rest of strong interactions, however, isn't confined to common processes with large cross sections such as the ``soft'' particle production, elastic scattering, or diffraction. It may well be that interesting, \textit{unusual} occurrences happen outside the framework of perturbative QCD---happen in some collective, or intrinsically nonperturbative, way. About fifteen years ago~\cite{Quigg:1994mp}, looking for an illustration of something unusual that might happen in the strong interactions, I appealed to a colleague to watch the event display at CDF and pick out a few \textit{atypical} events for me.\footnote{Run 1 of the Tevatron studied $\bar{p}p$ interactions at $\sqrt{s} = 1.8\tev$.}

\begin{figure}[tb]
\centerline{\includegraphics[width=0.8\textwidth]{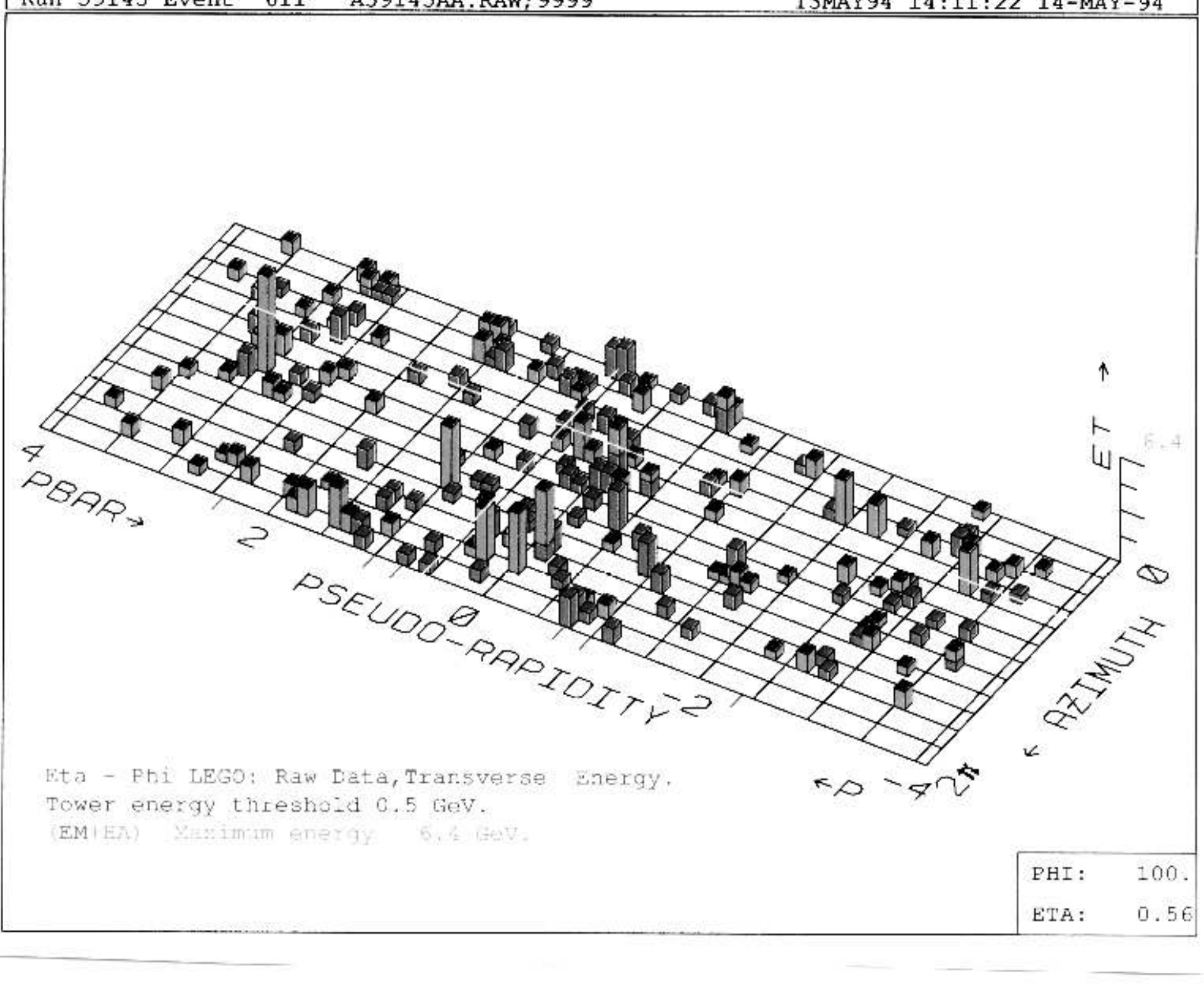}}
\caption{A hedgehog event captured in the CDF Run 1 detector. \label{fig:drasko}}
\end{figure}

The most interesting of these is shown in Figure~\ref{fig:drasko}, which comes from a $\sum E_\perp$ trigger, without any topological requirement. The LEGO$^{\scriptsize \textregistered}$ plot shows many bursts of energy:
More than a hundred active towers pass the display threshold of $0.5\gev$. The total transverse energy
in the event is $321\gev$, but it is not concentrated in a few sprays, it is everywhere. The central
tracking chamber records about sixty charged particles.

I am assured that this hedgehog event is authentic; it is not merely coherent noise in the counters. My colleague estimated such events to be about as common in the online event stream as $Z^0$ production and decay into
lepton pairs: about one in ten thousand triggers. I include this surprising picture as a reminder
that when we think about the strong interactions outside the realm of a single hard scattering,
we should think not only about the large diffractive and ``multiperipheral'' cross sections, but also about less common phenomena.

\begin{figure}[tb]
\centerline{\includegraphics[width=0.5\textwidth]{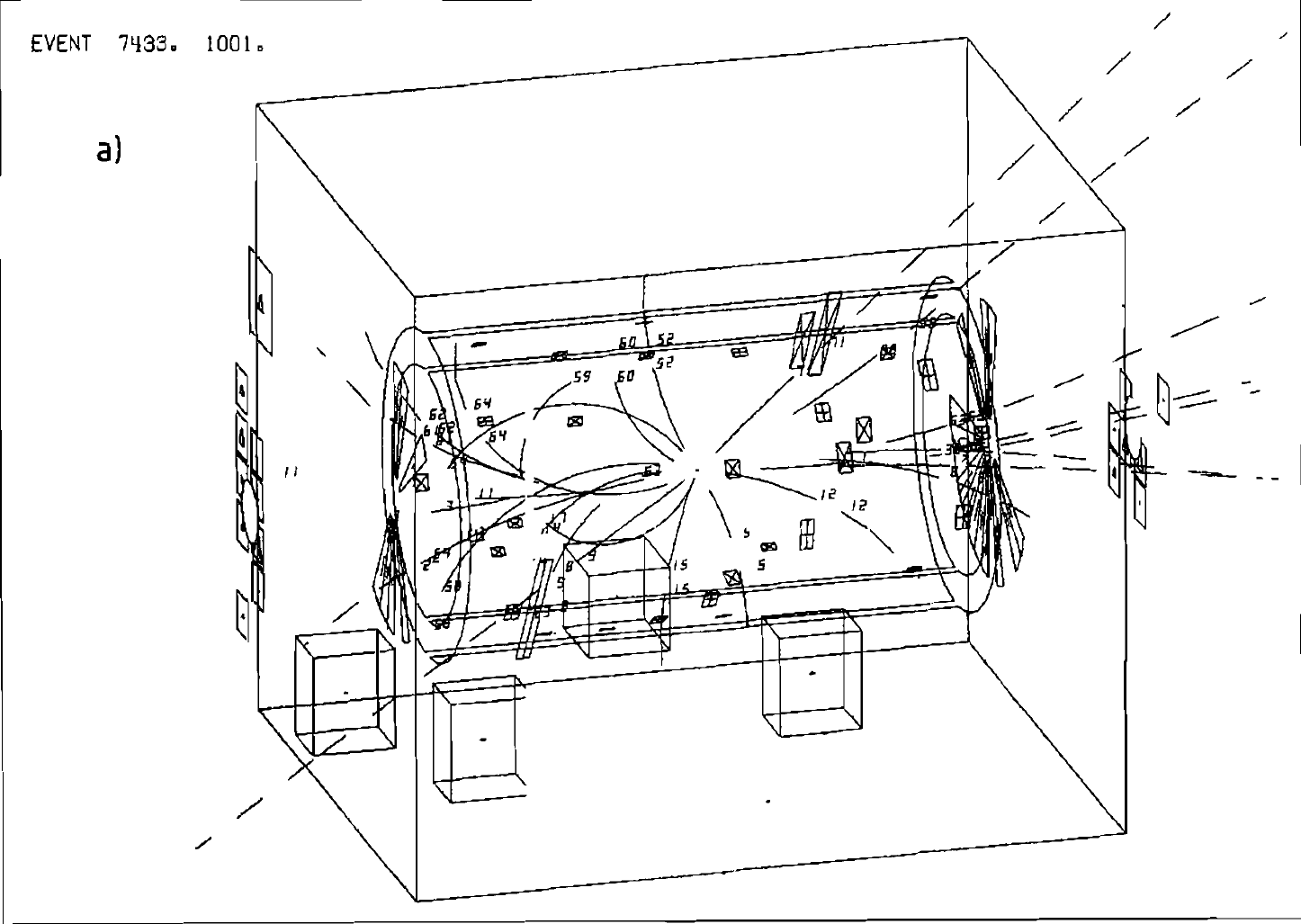}\includegraphics[width=0.5\textwidth]{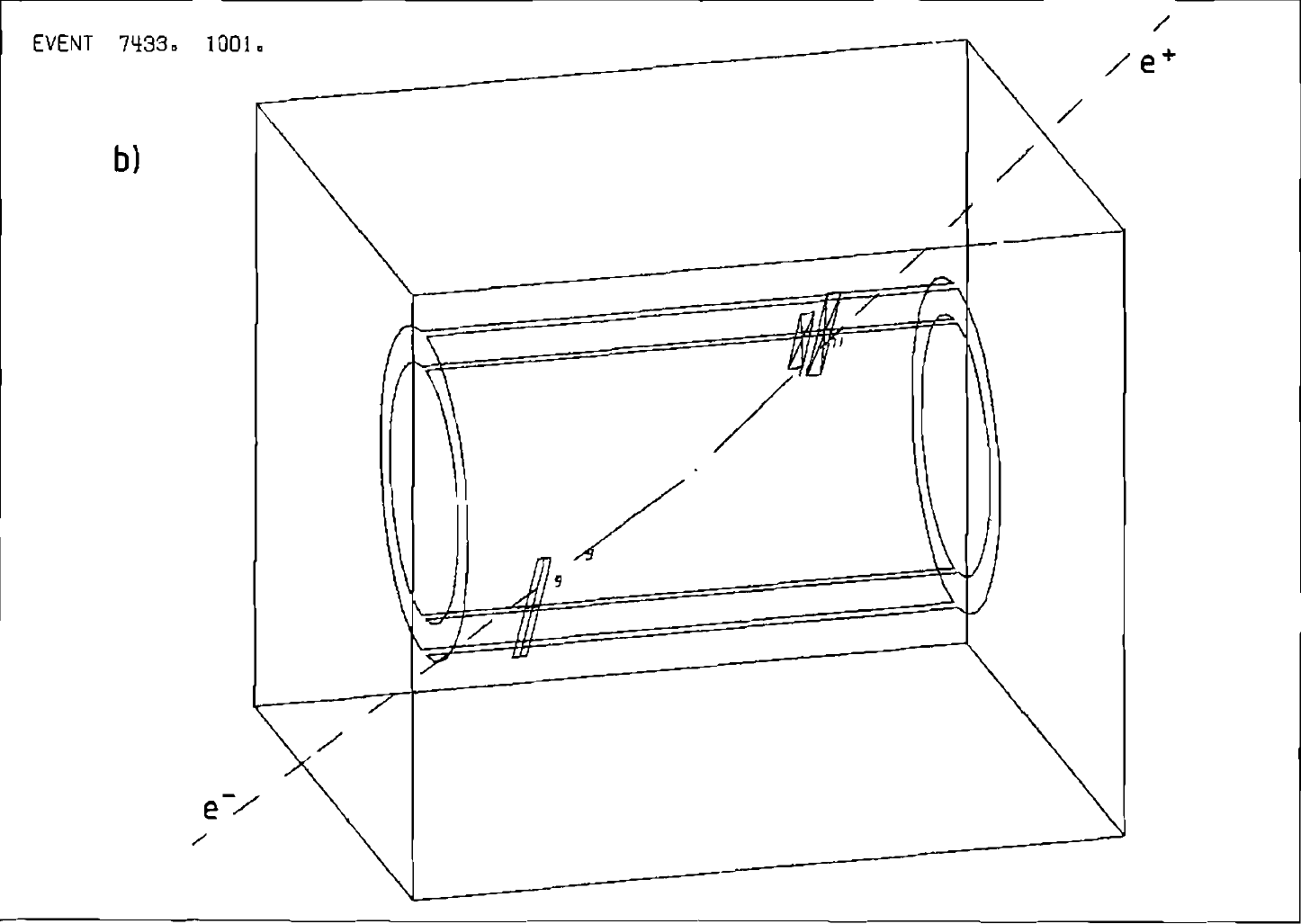}}
\caption{UA1 display of $Z^0 \to e^+ e^-$ event. (a) All reconstructed tracks and all calorimeter hits are displayed. (b) Same event, with display thresholds raised to $p_\perp > 2\gev$ for charged tracks and $E_\perp > 2\gev$ for calorimeter hits. Only the electron tracks survive these mild cuts. From~\cite{Arnison:1983mk}.\label{fig:ua1z}}
\end{figure}

Some thresholds are determined by detector performance, others are imposed by the exigencies of trigger management. When we prepare event specimens, raising \textit{display thresholds} is a powerful technique for removing clutter, the better to reveal the interesting hard-scattering action. An early example is shown in Figure~\ref{fig:ua1z}, 
which depicts one of the first $\bar{p}p \to Z^0 + \hbox{anything}$ candidates recorded by the UA1 experiment at CERN~\cite{Arnison:1983mk}, operating at $\sqrt{s} = 540\gev$. The left panel, which shows all reconstructed tracks and calorimeter hits, is a bit cluttered (by the standards of the early 1980s). The right panel, which shows only the tracks that satisfy $p_\perp > 2\gev$ and calorimeter hits with $E_\perp > 2\gev$, exhibits only the essential: stiff electron and positron tracks. Setting even a modest display threshold clears the weeds.
\textit{To look for physics in the weeds, it will pay to set the display thresholds as low as practicable.}

Bjorken has emphasized recently~\cite{BjWKT} that the usual parton-model language, in which we speak of individual, uncorrelated partons, is a convenient idealization that does not encompass the whole truth. This is self-evident when we consider static properties of hadrons, and long known for the spin structure of the nucleon as measured in polarized lepton-nucleon scattering in the limit $x_{\mathrm{Bj}} \to 1$~\cite{Hughes:1983kf}. Over the past decade, the study of generalized parton distributions~\cite{Ji:2004gf} has begun to explore simple correlations among partons. How the proton is composed---quarks bound by pairwise collisions, quark--diquark configurations, or three quarks linked through the three-gluon vertex---may be reflected in how the partons are distributed in impact-parameter space.\footnote{See~\cite{Shoshi:2002in} for an impact-parameter rendering of the proton, neglecting such correlations.} Such reasoning gives new encouragement to the search for novel event structures and strategies for interpreting them.

\centerline{\rule[-4pt]{12em}{1pt}}
The minimum-bias and lightly triggered data recorded during early LHC running will be valuable for developing intuition and for validating the assumptions that underlie searches for new physics in hard-scattering events. However, these data sets, to be gathered over steps in beam energy, also represent an important opportunity for exploration and discovery. One promising track will be to revisit the early studies of multiple production, which emphasized observables constructed from individual particles: topological cross sections (multiplicity distributions, including forward-backward asymmetries of multiplicity distributions), inclusive and semi-inclusive two-particle correlation functions, and charge-transfer studies---either between hemispheres or across a rapidity gap. For some classes of events, analyses of bulk properties, such as the studies of elliptic flow and determinations of thermodynamic parameters developed by the Relativistic Heavy-Ion Collider experiments at Brookhaven~\cite{Arsene:2004fa}, may be especially revealing.\footnote{It has happened before that a phenomenon established in nuclear physics has an analogue in particle physics under extreme conditions: consider quasielastic scattering from individual nucleons in lepton-nucleus scattering and the quark-parton picture of deeply inelastic scattering.}
We will need all the established methods---and more---to learn to see what the LHC data have to show.

Our goal in particle physics is to uncover the basic laws of nature, and to understand how those laws are expressed. The point of the exploratory spirit I advocate in this note is to be receptive to the plenitude of natural phenomena that our theories will have to explain---in reductionist or emergent terms, depending on the situation---in the expectation that the intrinsic fascination of new phenomena may guide us to the recognition of deeper principles.

\section*{Acknowledgments}
%\vspace*{12pt}
I thank Andrzej Buras and Gerhard Buchalla for warm hospitality at TUM and LMU, and gratefully acknowledge the generous support of the Alexander von Humboldt Foundation. I am grateful to Luis \'{A}lvarez-Gaum\'{e} and other members of the CERN Theory Group for their hospitality. I thank James Bjorken for stimulating discussions over many years, and for allowing me to post his recent ideas~\cite{BjWKT} for a wider audience. I am indebted to Drasko Jovanovi\v{c} for finding the event shown in Figure~\ref{fig:drasko} in the CDF online event stream. 
I thank Niccol\`{o} Moggi and William Wester for helpful information about zero-bias studies in the CDF Run 2 data set. 
Fermilab is operated by the Fermi Research Alliance under contract no.\  DE-AC02-07CH11359 with the U.S.\ Department of Energy.

\nonfrenchspacing

\end{document}